\documentclass[pdflatex,sn-mathphys]{sn-jnl}

\newcommand{\ket}[1]{\left\vert{#1}\right\rangle}

\jyear{2021}

\begin{document}

\title[Quantum error reduction with deep neural network]{Quantum error reduction with deep neural network applied at the post-processing stage}

\author*[1]{\fnm{Andrey} \sur{Zhukov}}\email{zugazoid@gmail.com}
\author[1,2,3]{\fnm{Walter} \sur{Pogosov}}\email{walter.pogosov@gmail.com}

\affil[1]{\orgname{Dukhov Research Institute of Automatics (VNIIA)}, \orgaddress{\city{Moscow}, \postcode{127055}, \country{Russia}}}
\affil[2]{\orgdiv{Institute for Theoretical and Applied Electrodynamics}, \orgname{Russian Academy of Sciences}, \orgaddress{\city{Moscow}, \postcode{125412}, \country{Russia}}}
\affil[3]{ \orgname{HSE University}, \orgaddress{\city{Moscow}, \postcode{109028}, \country{Russia}}}

\abstract{
Deep neural networks (DNN) can be applied at the post-processing stage for the improvement of the results of quantum computations on noisy intermediate-scale quantum (NISQ) processors. Here, we propose a method based on this idea, which is most suitable for digital quantum simulation characterized by the periodic structure of quantum circuits consisting of Trotter steps. A key ingredient of our approach is that it does not require any data from a classical simulator at the training stage. The network is trained to transform data obtained from quantum hardware with artificially increased Trotter steps number (noise level) towards the data obtained without such an increase. The additional Trotter steps are fictitious, i.e., they contain negligibly small rotations and, in the absence of hardware imperfections, reduce essentially to the identity gates. This preserves, at the training stage, information about relevant quantum circuit features. Two particular examples are considered that are the dynamics of the transverse-field Ising chain and XY spin chain, which were implemented on two real five-qubit IBM Q processors. A significant error reduction is demonstrated as a result of the DNN application that allows us to effectively increase quantum circuit depth in terms of Trotter steps.
}



\maketitle

\section{Introduction}\label{sec:intro}

Quantum information is a fast developing field that aims to utilize quantum properties, such as quantum interference and entanglement \cite{horodecki2009quantum}, to perform functions related to computing \cite{kandala2017hardware, barends2015digital}, communication \cite{mattle1996dense,zhukov2019quantum}, and simulation \cite{georgescu2014quantum}.
State-of-the-art quantum computers are already capable of solving many problems \cite{peruzzo2014variational, o2016scalable, mohseni2017commercialize, li2017efficient,preskill2018quantum}, which, however, are not of practical importance yet, because of relatively high quantum hardware error rates. Particularly, such processors can be useful for solving evolutionary problems, such as investigation of the quantum dynamics of the Fermi-Hubbard model, see, e.g., Ref. \cite{arute2020observation}. However, the simulation of the dynamics of such systems at long times requires a large number of Trotter decomposition steps of evolution operator. This leads to the fact that a large number of quantum gates are required for simulation, which means that the outcomes from the quantum computer become too noisy~\cite{zhukov2018algorithmic,we2020}.

An attempt to enhance capabilities of quantum devices, from one hand, and machine learning methods, from another hand, has led to the merger of these areas \cite{biamonte2017quantum, Dunjko_2018, perdomo2018opportunities, ciliberto2018quantum, schuld2015introduction}, which gave rise to a new discipline known as quantum machine learning (QML). For example, hybrid quantum-classical systems were proposed \cite{Benedetti_2019}, where QML algorithms work with both classical data from quantum devices and quantum algorithms on a quantum processor. Thus using machine learning \cite{nielsen2015neural, carleo2019machine} to reduce the burden of classical information processing for quantum computation results has recently become an area of intense interest. The development of classical machine learning technologies can be useful for such tasks as the verification of quantum devices \cite{lennon2019efficiently}, quantum error correction \cite{nautrup2019optimizing,Baireuther2018machinelearning, Andreasson2019quantumerror}, quantum control \cite{kalantre2019machine, vozhakov2021state, bukov2018reinforcement,niu2019universal}, quantum state classification \cite{babukhin2019nondestructive, carrasquilla2017machine}, and quantum state tomography \cite{altepeter2005photonic, torlai2018neural,neugebauer2020neural,lohani2020machine, sehayek2019learnability}.
For example, using deep machine learning in quantum tomography, it became possible to reduce errors in the preparation and measurement of quantum states~\cite{Straupe19} or predict if the set of measurements is informationally complete to uniquely reconstruct any given quantum state~\cite{Straupe2021}.

The main difficulty associated with neural networks in the context of noisy quantum machines is that it is necessary to have an access to the "ideal" data for their training. In principle, such data can be obtained from the classical simulation, but in the case of a large number of qubits, this becomes impossible. Methods that can be utilized to overcome this problem ~\cite{czarnik2020error,strikis2020learning,kim2020quantum} include the usage of quantum circuits largely composed of Clifford gates that can be simulated classically. In our work, we address the problem of digital simulation of quantum dynamics and propose another method for obtaining quasi-ideal data for training using only a quantum computer, without a classical simulation. Our method also respects a periodic quantum circuit structure. As a source of quasi-ideal data, we consider outcomes from real quantum computer, which correspond to a limited number of Trotter steps. At the training stage, circuit depth is artificially increased by incorporation of certain number of "empty" Trotter steps, which contain zero or negligible rotation angles and thus reduce to identity gates for the ideal execution of quantum gates. The neural network is trained to transform such data with increased noise level towards their quasi-ideal counterpart. After being trained, the network is applied to improve new noisy data, which are obtained by increasing Trotter number, while none of Trotter blocks is fictitious. We demonstrate this rather generic method by studying the time evolution of the magnetization of the transverse-field Ising spin chain and XY spin chain using IBM Q Athens and Bogota quantum processors, respectively. Both processors consist of five qubits and have a linear connectivity topology that matches spin chain topology.

We believe that our approach can be of particular interest in the context of digital simulation of different quantum systems. Indeed, such a simulation generally suffers from error accumulation problems, so that a very limited number of systems have been simulated so far. For example, Ref. \cite{arute2020observation} reports on simulation of Fermi-Hubbard model in a one-dimensional geometry using state-of-the-art superconducting quantum processor, while much more interesting simulation of the same model in two dimensions remains challenging. In order to reduce errors, several techniques have been used \cite{arute2020observation}, which allowed to increase Trotter step number and thus to trace the system's dynamics at a longer time. The application of neural networks at the post-processing stage is a prospective tool that can be used in combination with other techniques.

The article is organized as follows. Section II deals with the formulation of the problem of the dynamics of the Ising spin chain in a transverse field. Section III details our method for reducing hardware errors using deep machine learning. Section IV describes the practical implementation of the method on a real processor for the transverse-field Ising model. In Section V we apply our method to another system, which is XY spin chain. Section VI presents the Conclusions.

\section{Trotterization of quantum dynamics}

Our approach can be applied for the simulation of various quantum many-body systems, but in the beginning we are going to illustrate it in a detail with a paradigmatic example of a system that is widely used for benchmarking state-of-the-art quantum machines and their ability to work as digital, analog, or digital-analog \cite{we2020} quantum simulators. This is the transverse-field Ising model, for which we choose a one-dimensional geometry because it perfectly matches the topology of a five-qubit IBM Q Athens processor, which has one of the lowest gate error rates among state-of-the-art quantum processors, as well as one of the largest quantum volumes and therefore it is suitable for the demonstration of our concept, which requires quantum circuits of a rather large depth. In Section V we consider another example, which is XY spin chain. The transverse-field Ising Hamiltonian reads as
\begin{equation}
    H =  -\sum_{j}h_{j}\sigma^{x}_{j} -\sum_{<ij>}J_{ij}\sigma^{z}_{i}\sigma^{z}_{j},
    \label{eq:ham}
\end{equation}
where $\sigma_j$ are Pauli operators acting in the space of $j$'s spin, represented directly by $j$'s qubit of the chain, $h_{j}$ are local fields, and $J_{ij}$ are coupling constants which are supposed to be nonzero only for nearest neighbors.

One of the standard approaches to simulate quantum systems on digital quantum processor is based on Trotter decomposition of the evolution operator. We are going to concentrate on the ndynamics of magnetic characteristics. In the case of the Hamiltonian $H$ consisting of two non commuting contributions, $H=H_A + H_B$, Trotter decomposition transforms the evolution operator for $H$ into the sequence of evolution operators for $H_A$ and $H_B$
\begin{equation}
      e^{-it(H_{A}+H_{B})} \approx (e^{-iH_{A}\frac{t}{N}}e^{-iH_{B}\frac{t}{N}})^{N},
      \label{eq:trotter}
\end{equation}
which is asymptotically exact in the limit of large Trotter step number $N \longrightarrow \infty$, otherwise Trotterization produces digitization (discretization) error. For the transverse-field Ising Hamiltonian, the appropriate splitting of the full Hamiltonian is
\begin{equation}
    H_{A} = -\sum_{j}h_{j}\sigma^{x}_{j},
\end{equation}
\begin{equation}
    H_{B} = -\sum_{<ij>}J_{ij}\sigma^{z}_{i}\sigma^{z}_{j},
\end{equation}
so that the evolution operators for both $H_A$ and $H_B$ can be further split exactly into evolution operators for each $-h_{j}\sigma^{x}_{j}$ and $-J_{ij}\sigma^{z}_{i}\sigma^{z}_{j}$ terms, since these contributions do commute to each other. All these operators can be implemented in real quantum computers in a standard way using quantum circuits language.

\begin{figure}[t]
\begin{center}
\includegraphics[width=0.50\textwidth]{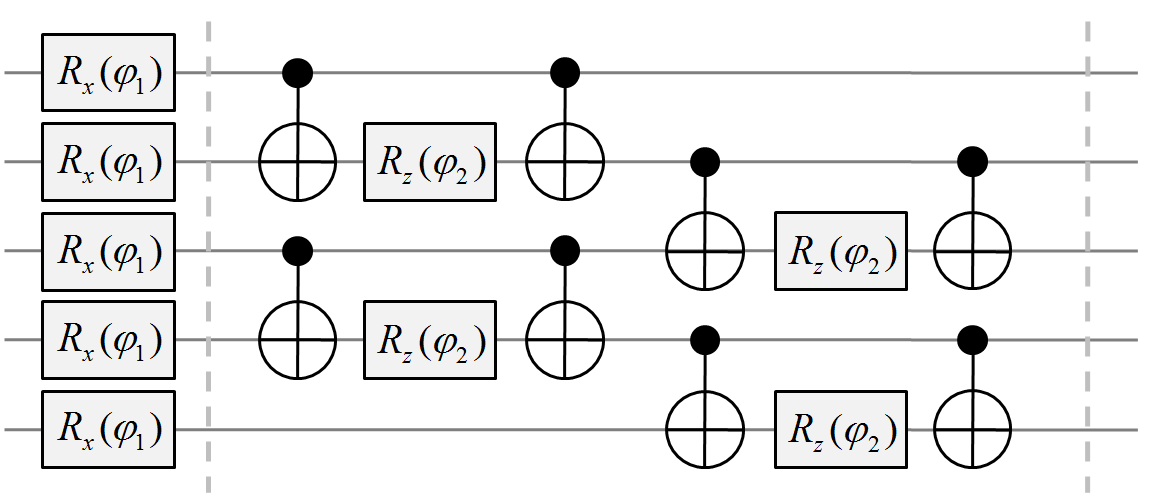}
\caption{\label{fig:circ} Quantum circuit for 5-spin chain: the structure of the single Trotter step.}
\end{center}
\end{figure}

Let us concentrate on the quantum circuit for $N_q = 5$ transverse-field Ising chain. We will present our results for the case, when all coupling constants and free fields for spins of the chain are equal to each other, $J_{ij} = J = 2$, $h_j=h=1$. Our quantum circuit corresponding to the single Trotter step for the time interval $\delta t = t/N$ is shown in Fig. \ref{fig:circ}, where single-qubit rotations are defined as
$R_x(\varphi_1)=e^{-i \frac{\varphi_1}{2} X}$ and $R_z(\varphi_2)=e^{-i \frac{\varphi_2}{2}Z}$, where $\varphi_1=2h \,\delta t$ and $\varphi_2=2J \delta t$. Hereafter, $X$, $Y$, and $Z$ refer to Pauli gates.

\section{Error reduction with DNN}

\begin{figure}
\begin{center}
\includegraphics[width=0.750\textwidth]{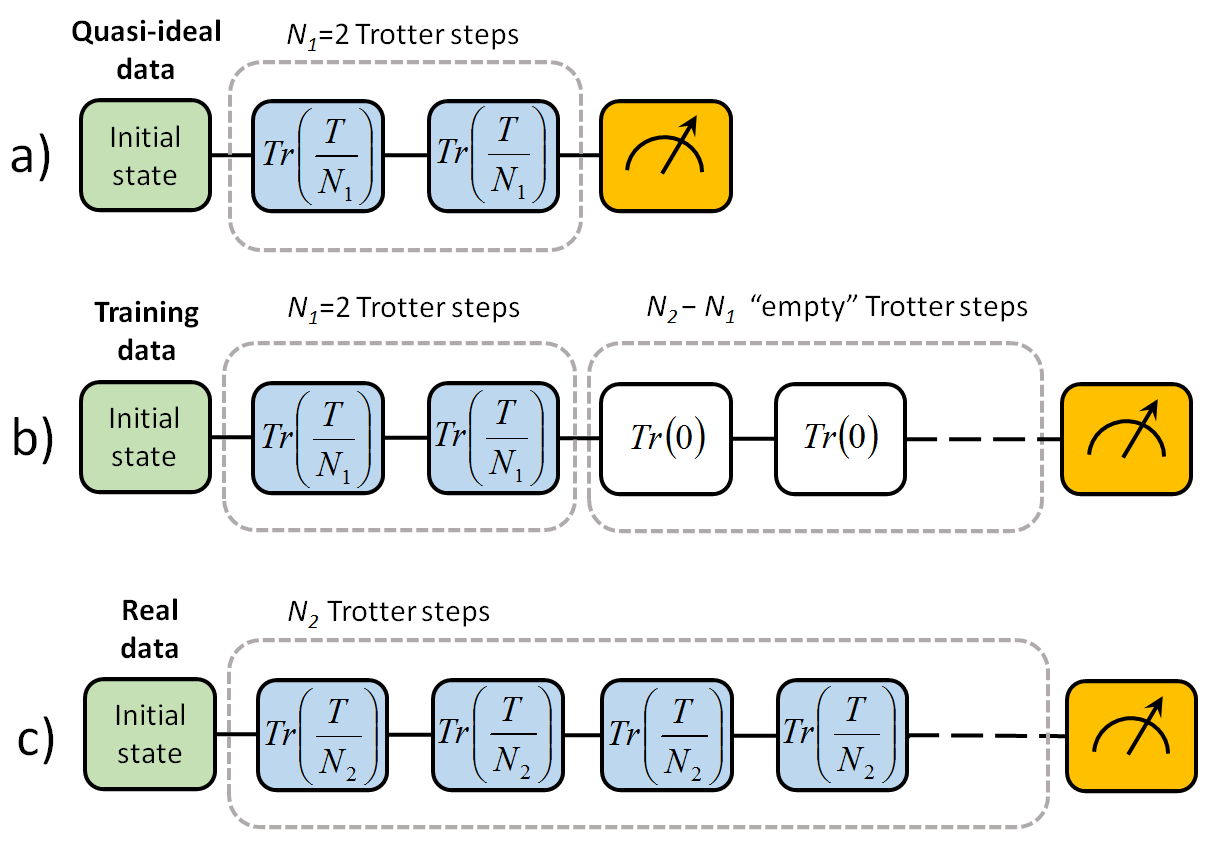}
\caption{\label{fig:trotters}Schematic view of our approach. It consists of three steps: generation of quasi-ideal data with relatively shallow circuit (a); training the DNN -- the data with artificially increased Trotter steps number are transformed towards quasi-ideal data (b); the trained DNN is applied to raw experimental data with the same Trotter step number as at the second stage (c).}
\end{center}
\end{figure}

First of all, neural network needs to be trained. Imagine that we have reliable, i.e., noise-free data corresponding to the outcomes from an ideal quantum computer. They can be represented in a classical form by some characteristics of a quantum state, i.e., quantum mean values or correlators, that does not require a full quantum tomography. Similar data but affected by hardware imperfections can be obtained as outcomes from a noisy quantum computer. Next, the neural network can be trained to transform noisy data towards ideal results. After being trained, the network can be applied to other noisy data, which were not included in the initial training data set, but were affected by nearly the same amount of noise. The most attractive potential application of this rather general idea is an effective increase of an accessible quantum circuit length (for example, increase of  Trotter step number) due to the improvement of data quality at the post-processing stage.

This general approach requires an "ideal" dataset to train the neural network. In principle, such a dataset can be obtained using a classical simulator. However, in the case of large quantum processors with many qubits, classical simulation will be possible only in some special cases. Thus, it might seem that an absence of an ideal training dataset makes this idea useless for real applications using quantum machines with many qubits.

Let us, however, assume that we have a reliable enough low-noise quantum computer. Let our task be to analyze the digitized evolution of the quantum system for some time interval $(0, T)$ taking into account as large number of Trotter steps as possible. To this end, we will proceed as follows. Let us execute the algorithm for $N_1$ Trotter steps for the total evolution time $T$, as shown schematically in Fig. \ref{fig:trotters} (a) for $N_1=2$, where $Tr(\delta t)$ stands for the evolution operator consisting of a single Trotter step of length $\delta t$. Each Trotter step has a length of $T/N_1$ at this stage. The Trotter steps number $N_1$ must be chosen in such a way as to keep the total gates error small. Such data generated by the quantum computer can be considered as quasi-ideal in the sense that errors due to hardware imperfections are small.  Note that, in practical situations, optimal $N_1$ can be estimated from the balance between total gate error, which increases as circuit depth grows, and Trotterization error, which decreases as circuit depth grows. Optimal $N_1$ thus can be found from the condition of minimum of the total error.

Next, let us consider the method of training of the neural network using quasi-ideal data. Imagine that our goal is to increase the total Trotter step number from $N_1$ to $N_2 = c N_1 $  for the same time interval $(0,T)$, where $c$ is a natural number. Apparently, the level of noise will be increased in this case and this must be taken into account at the training stage. The training can be performed as follows. We extend the quantum circuit for $N_1$ Trotter steps and the total evolution time $T$ by adding $N_2-N_1$ "empty" Trotter steps, as illustrated in Fig. \ref{fig:trotters} (b). These fictitious Trotter steps must include rotations on zero or negligibly small angles making each ideal "empty" Trotter block essentially equivalent to the identity gate. For the perfect execution of the quantum algorithm, the new data must coincide with the quasi-ideal data, but due to the increased number of noisy quantum gates, they do differ. The neural network is to be trained to transform these noisy data towards quasi-ideal data. After the training, we can run the quantum computer for $N_2$ Trotter steps and total evolution time $T$, such that each Trotter step has a length $T/N_2$, see Fig. \ref{fig:trotters}(c). The quantum gate number, as well as the structure of the quantum circuit and its composition in terms of quantum gates are similar to the case with $N_1$ "real" and $N_2-N_1$ "empty" Trotter steps for which the neural network has been trained ($N_2$ Trotter blocks in total in both cases). Let us stress that it is known that circuit structure is of importance since it directly influences quantum computer performance  \cite{proctor2020measuring}. Finally, at the post-processing stage, we apply the trained network for these last data, as illustrated in Fig. \ref{fig:trotters} (c), that allows to partially get rid of noise effects. There are certain analogies between our approach and zero-noise extrapolation technique \cite{Bravyi}.

Note that there can be different permutations between positions of "real" and "empty" Trotter blocks in the quantum circuit at the training stage. This leads to the variety of different schemes, the efficiency of which is worth to be studied. Also note that our approach does not use ideal data at the training stage, and therefore, it is hard to expect that the post-processing based on the DNN application would demonstrate higher accuracy compared to quasi-ideal data with $N_1$ Trotter steps, which are influenced by gate errors.

\section{Magnetic properties of transverse-field Ising chain}

\begin{figure}[t]
\begin{center}
\includegraphics[width=0.50\textwidth]{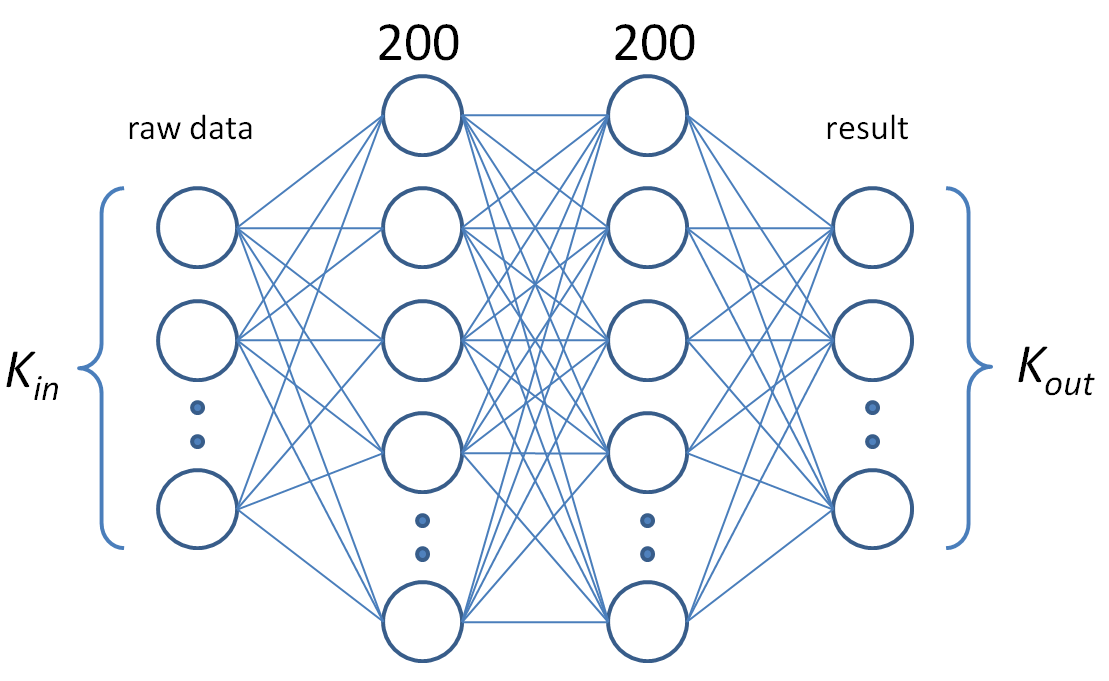} 
\caption{\label{fig:neuro0} The feed-forward DNN architecture used in our problems for improving simulation results. Input layer consists of $K_{in}$ neurons. Output layer contains $K_{out}$ neurons. Two hidden layers consist of 200 neurons both. We use sigmoid activation function after both hidden and output layers. During training, the neural network changes its internal parameters in order to achieve a better mapping of the input noisy data to the quasi-ideal counterpart.}
\end{center}
\end{figure}

Now we apply our idea for the digital quantum simulation of 5-spin transverse-field Ising chain using IBM Q Athens quantum processor. We provide an illustrative example of a neural network, which is trained to improve noisy data for the dynamics of magnetization $\langle Z_j \rangle$ of each spin, $j = 1 \ldots 5$, in $z$ direction. We utilize a simple feed-forward deep neural network whose architecture is shown in Fig.~\ref{fig:neuro0}. The number of input neurons is chosen to be $K_{in}=15$, which corresponds to the magnetization of each spin in three directions $x$, $y$, and $z$. DNN has two hidden layers consisting of 200 neurons both. The output is a 5-dimensional layer representing magnetization of each spin in the $z$ direction only since we have chosen magnetization in this direction as a target quantity. Note that, of course, the number of input and output neurons can be different from the case considered here since different characteristics of the spin system can be used in datasets, so the neural network shown in Fig. \ref{fig:neuro0} is just a specific example. In Section V a similar network will be used for XY spin chain, but with $K_{in}=K_{out}=5$ that corresponds to magnetizations only in $z$ direction, since our method will be supplemented by the post-selection of outcomes according to the excitation-number conservation condition.

For a given pair $(N_{1},c)$, a set of curves for $\langle X_j \rangle$, $\langle Y_j \rangle$, $\langle Z_j \rangle$ as functions of time is obtained from real quantum computer for all possible initial conditions, which correspond to the computational basis (such as $\ket{01101}$, $\ket{10110}$, etc.). The same DNN is thus trained for all these initial conditions and it is able to improve raw data for all of them. The number of such initial states is $2^{N_{q}}$, where $N_{q}$ is the number of the qubits ($N_{q}=5$ in our example). As a whole, for a given Trotter step number, the quantum computer generates a quasi-ideal dataset of the size $D=2^{N_{q}}\times N_{q} \times K \times B$, where $K$ is the number of points along a given time interval $(0,T)$, $B=3$ is the number of characteristics for each spin we are interested in (magnetization in $x$, $y$, and $z$ directions in our example). As usual in the IBM Q system, 8192 measurements for each point are performed. For our DNNs, we tested different activation functions such as the sigmoid or ReLu and found that the sigmoid function yielded the best results.

The quality of data improvement is estimated by the comparison between quasi-ideal data and data improved by DNN for the whole time interval $T$. The quality criterion for DNN is the Mean Square Error (MSE), defined as
\begin{eqnarray}
E_{A,B}=\frac{1}{2D} \sum_{i,j,k,l} \left(  \langle M_k^{j,l} (t_i) \rangle_{A}- \langle M_k^{j,l} \rangle_{B}  \right)^2,
\label{MSE}
\end{eqnarray}
which depends on difference between training dataset ($A$) and quasi-ideal data ($B$) . Here $M_k^{j,l} (t_i)$ is the value of magnetization of spin $j$ along axes $k=x, y, z$ for time $t_i$, and initial state $l$. Training is performed by minimization the MSE. For this, we applied the Adam algorithm \cite{adam}. The network was trained during 50000 epochs. Network parameters were saved for every 100 epochs in order to find optimal training time. The networks are trained separately for each couple of parameters $N_1$ and $c$. In our examples, we focus on $N_1=2$ and $c=$2, 3, 4 (three DNNs in total). All data for training and testing the neural network were obtained using the real quantum processor.  In our illustrative examples, the value of $T=2 / J$ is taken to be several times larger than the typical maximum time for which the approximation with $N_1=2$ has a negligibly small digitization error. Of course, such a small value of $N_1$ as $N_1=2$ is meaningful only for demonstration purposes. An increase of $N_1$ leads to the enhancement of hardware error rates which become too high for our demonstration. Three DNNs are trained in total, each corresponding to a given $c$.

\begin{figure}
\begin{center}
\includegraphics[width=0.50\textwidth]{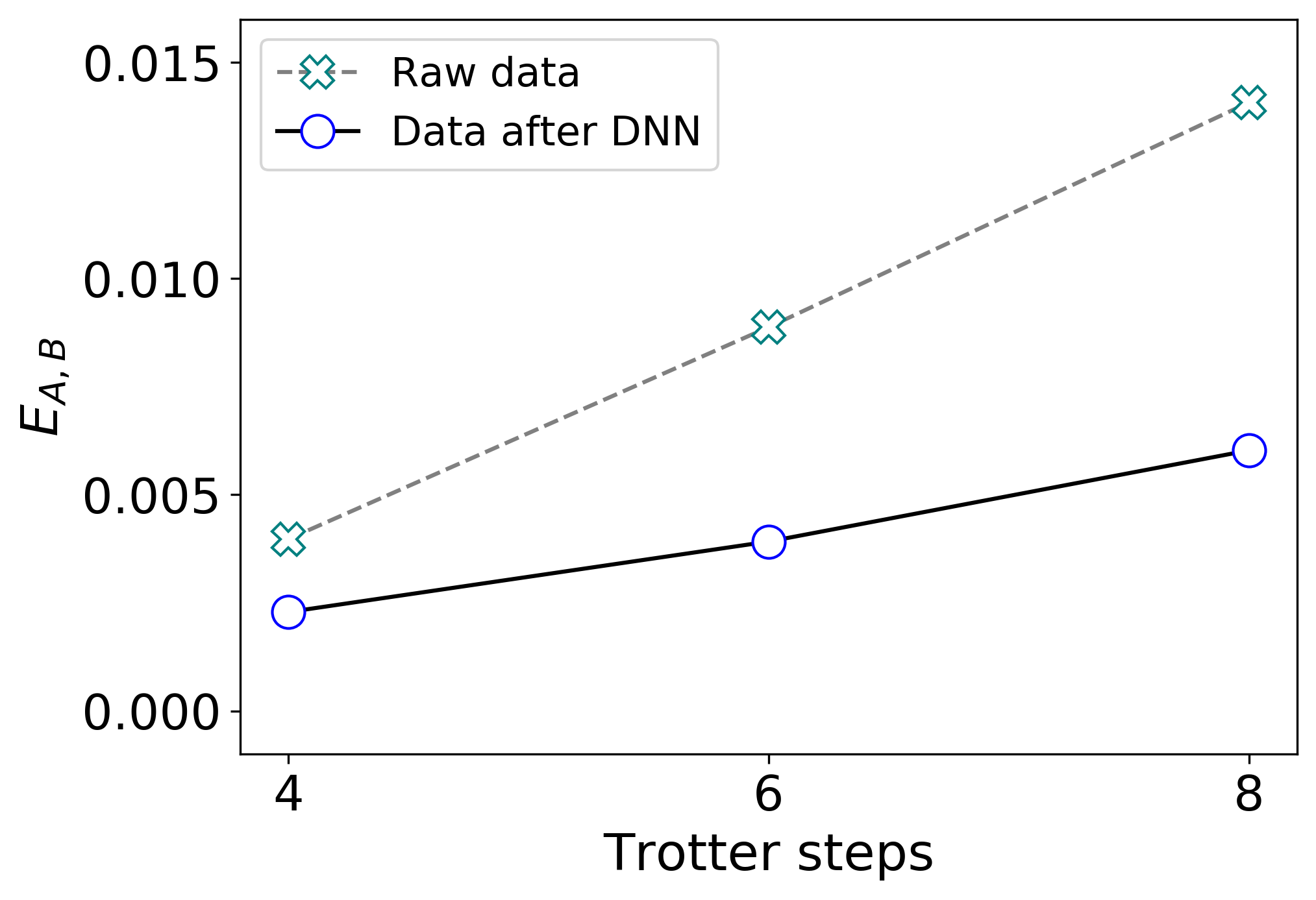} 
\caption{\label{fig:fig_tr2_e}MSE between ideal simulation data for a given Trotter step number and experimental data from an IBM Athens 5-qubit quantum processor improved by the network (o-shape symbols and solid line) as well as raw data (x-shape symbols and dashed line). The errors were averaged over time and all initial states from the computational basis.}
\end{center}
\end{figure}

\begin{figure}
\begin{center}
\includegraphics[width=0.50\textwidth]{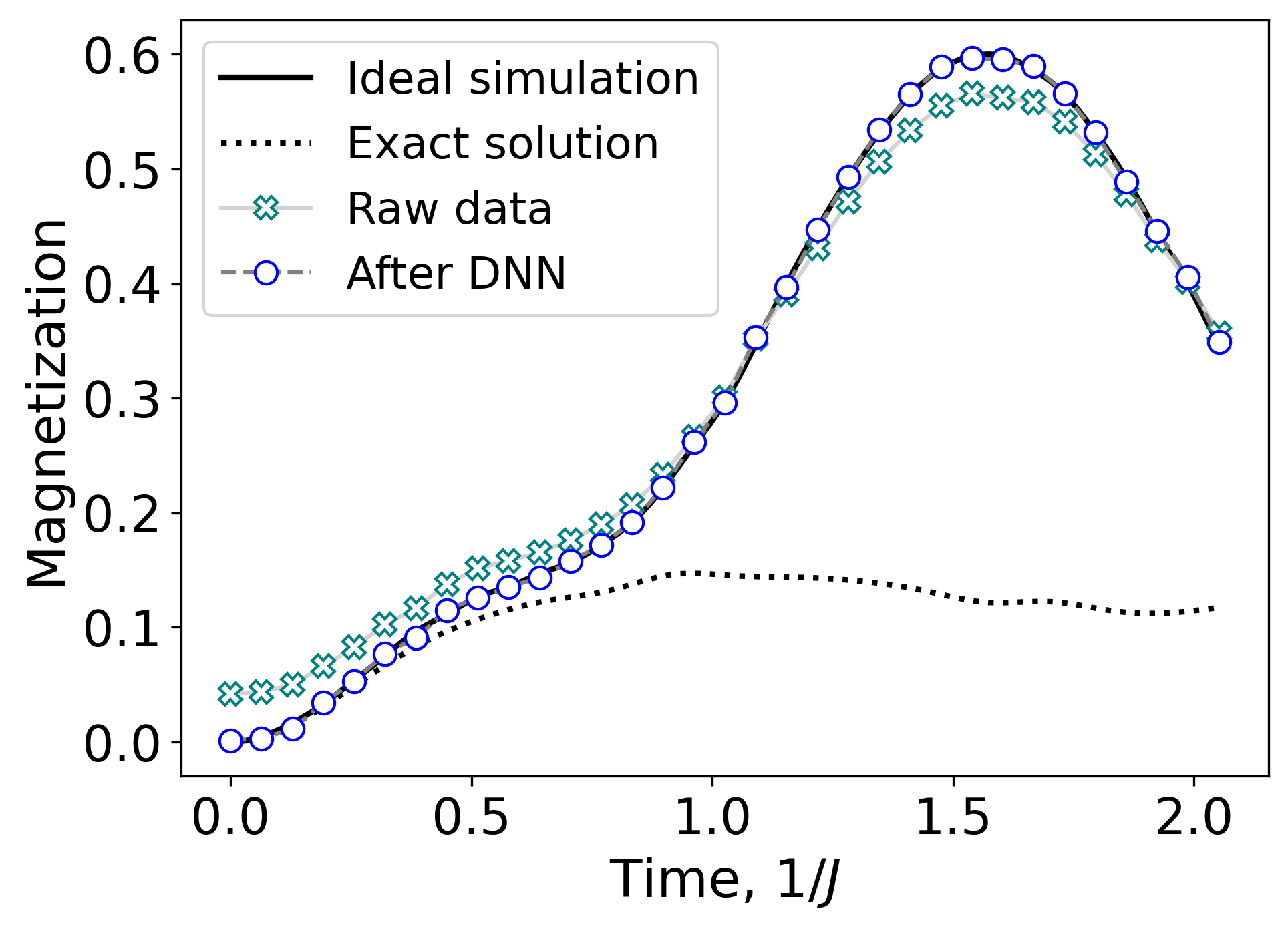} 
\caption{\label{fig:fig_tr12}
The mean spin chain magnetization along the $Z$ axis as a function of time for the initial state $\ket{00000}$. Solid curves are the results of ideal simulations for given Trotter step number $N_1=2$. Raw experimental data for magnetization (x-shape symbols) were obtained on the real 5-qubit quantum processor IBM Athens. The results of applying the neural network are shown with o-shape symbols. The neural network was trained using ideal data from the simulator (see the text). The dotted line shows the reference ideal evolution of magnetization free of Trotterization errors. For all curves $J=2$, $h=1$.}
\end{center}
\end{figure}

Our results for errors $E_{A,B}$ at $k=z$ are presented in Fig. \ref{fig:fig_tr2_e}. This figure shows the MSE (\ref{MSE}) between both raw data and data improved by DNN (A) and exact data for the same Trotter step number (B). Averaging over all initial conditions is performed. It is seen from this figure that the improvement due to the neural network is significant generally decreasing MSE in several times. In order to demonstrate the ability of the DNN to improve the dynamics, in Fig. \ref{fig:fig_tr12} we show the the mean magnetization of 5 spins $\langle \overline{Z_j} \rangle$ averaged over spins $j$ as a function of time for the initial state $\ket{00000}$. It presents the results of improvement for quasi-ideal data corresponding to $N_1=2$. For illustrative purposes, in this particular case, DNN was trained using ideal data from the simulator with the same Trotter step number. This provides an important test -- Fig. \ref{fig:fig_tr12} shows that the DNN is able to transform raw data essentially to the ideal counterpart. The figure shows four curves: the result of the ideal simulation with two Trotter steps (solid curve), the exact solution corresponding to the Schrodinger equation in the absence of Trotterization errors (dotted curve), raw data (x-shaped symbols), and data improved by DNN (circle-shaped symbols). It reveals a nearly perfect matching between ideal simulation and data improved by DNN. A similar level of improvement has been achieved for all other initial conditions considered.

\begin{figure}
\begin{center}
\includegraphics[width=0.99\textwidth]{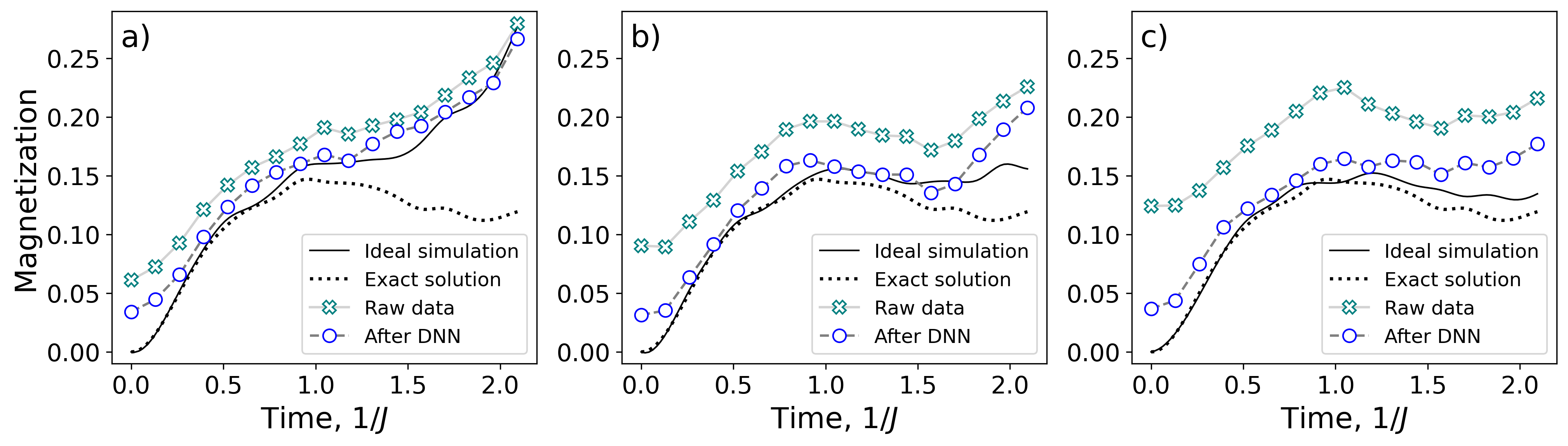}
\caption{\label{fig:fig_tr2_ev}
The mean spin chain magnetization along the $Z$ axis as a function of time for the initial state $\ket{00000}$. Solid curves are the results of ideal simulations for given Trotter steps number $N_2$:  4 (a), 6 (b), 8 (c). Experimental data for magnetization (x-shape symbols) were obtained on the real 5-qubit quantum processor IBM Athens. The results of applying the neural network are shown with o-shape symbols. For each case, the neural network was trained using experimental data for $N_1=2$ and $c=2,3,4$, respectively (see the text). The dotted line shows the exact reference evolution of magnetization free of Trotterization errors. In all cases $J=2$, $h=1$.}
\end{center}
\end{figure}

\begin{figure}
\begin{center}
\includegraphics[width=0.99\textwidth]{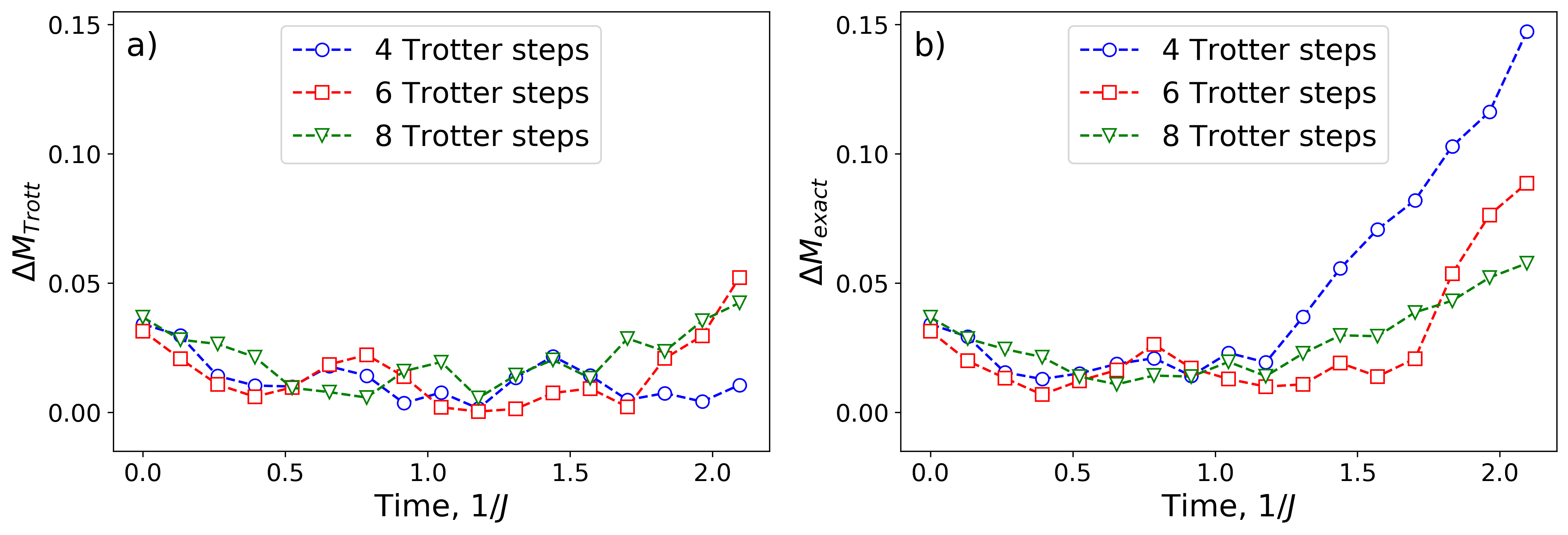} 
\caption{\label{fig:fig_errors}
Errors after DNN application as a function of time for the initial state $\ket{00000}$ -  (a) difference between improved results and ideal simulation for the same Trotter step number;  (b) difference between improved results and exact results free of Trotterization error.}
\end{center}
\end{figure}

Next, the improved results for $N_1=2$ and $c=2,3,4$ are compared to the ideal results free of hardware errors, which include the results of the full solution of the time-dependent Schrodinger equation corresponding to the infinite Trotter steps numbers. In Fig. \ref{fig:fig_tr2_ev} we plot such time dependencies of $\langle \overline{Z_j} \rangle$ averaged over spins $j$, which are referred to as "exact results", for the initial state $\ket{00000}$ (dotted lines). Also shown are the ideal results for a given Trotter step number (solid line), referred to as "ideal simulation". raw results (x-shaped symbols) as well as the improved results (o-shaped symbols), which are plotted at three values of $c$: 2 (a), 3 (b), 4 (c). Fig. \ref{fig:fig_errors} shows deviations between the results improved $\langle \overline{Z_j} \rangle_{DNN}$ and ideal results with the same Trotter step number $\langle \overline{Z_j} \rangle_{Trott}$, $\Delta M_{Trott} = \langle \overline{Z_j} \rangle_{DNN} - \langle \overline{Z_j} \rangle_{Trott}$ (a), as well as between the results of the full solution of Schrodinger equation $\langle \overline{Z_j} \rangle_{exact}$, $\Delta M_{exact} = \langle \overline{Z_j} \rangle_{DNN} - \langle \overline{Z_j} \rangle_{exact}$ (b).

In general, it is seen from Figs. \ref{fig:fig_tr2_ev} (c) and \ref{fig:fig_errors} (b) that DNN brings the raw results closer to the ideal curves for all values of $c$ we considered. For example, raw data for $N_1=N_2=2$, see Fig. \ref{fig:fig_tr12},  were reasonably close to the exact results free from both digitization and hardware errors up to the maximum time $t \lesssim 0.5$. The DNN extended this range to $t \lesssim 1.0$ for $N_2=4$ and $t \lesssim 1.5$ for $N_2=6$. The extension of the time interval, when there exists a good agreement between the results of the full solution of the time-dependent Schrodiger equation and the outcomes from the real quantum computer enhanced by DNN, is one of the key results of our proof-of-principle demonstration.

Let us also note that our approach typically leads to the spurious increase of the error at short times. We believe that this artifact is not crucial since quasi-ideal data turn out to be accurate in this range of parameters so that they can be treated as reference results generated by the quantum computer.

More advanced DNN architectures together with different types of input characteristics of the quantum state as well as the incorporation of other error mitigation techniques can further enhance the abilities of DNN at the post-processing stage. This requires an interplay between the physics of the underlying model we simulate and DNN architecture. Another improvement can be due to the more advanced Trotterization that accounts for higher orders of Trotter step length.

Note that we expect that, in general, for the successful training of a neural network, there is no need for data corresponding to the full set of initial states of the computational basis. All that is needed is the information about the dynamics for various initial states that correspond to different patterns of the system's dynamics. Therefore, as it is expected, there will be no need to include all $2^ {N_q}$ possible initial states to the training set that is crucial for large $N_q$. It is possible to apply, for example, Monte-Carlo methods when choosing an initial state and stop training the neural network until the level of errors becomes acceptable, or the addition of new data ceases to improve results. 


\section{Magnetic properties of XY spin chain}

\begin{figure}[t]
\begin{center}
\includegraphics[width=0.60\textwidth]{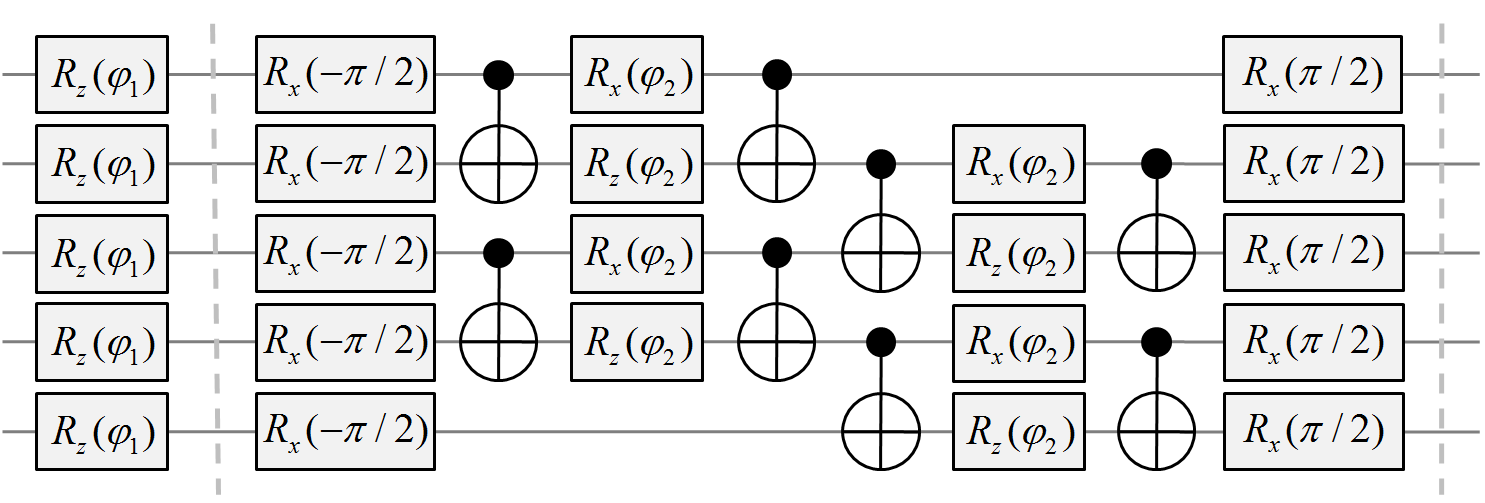} 
\caption{\label{fig:circ_xy} Quantum circuit for 5-spin XY chain: the structure of the single Trotter step.}
\end{center}
\end{figure}

In this section we consider another quantum system, which is XY quantum chain and apply our method for this particular case. The system considered is described by the Hamiltonian
\begin{equation}
    H =  -h\sum_{j}\sigma^{z}_{j} -J\sum_{<ij>}(\sigma^{x}_{i}\sigma^{x}_{j}+\sigma^{y}_{i}\sigma^{y}_{j}),
    \label{eq:XYham}
\end{equation}
where the interaction acts between nearest neighbors in the chain. We concentrate on the chain consisting on five spins, which is simulated using five-qubit quantum IBM Q processor Bogota with linear connectivity. Quantum circuit for this chain corresponding to the structure of the single Trotter step is shown in Fig.~\ref{fig:circ_xy},
where $\varphi_1=2h \,\delta t$ and $\varphi_2=2J \delta t$.
We apply the same method by improving the results of the quantum simulation using the neural network at the post-processing stage, as illustrated in Figs. \ref{fig:trotters} and \ref{fig:neuro0}.

As a dataset for the neural network, we use magnetizations in $z$ direction only, $\langle Z_j \rangle$. Thus, the parameters of the neural network, shown in Fig. \ref{fig:neuro0}, are $K_{in}=K_{out}=5$. The Hamiltonian conserves the excitation number and this allows us to use a post-selection as an additional tool to mitigate errors: at the training stage, as a quasi-ideal data we use the experimental data for $N_1=2$, for which we discard the data with wrong excitation numbers. This allows us to noticeably improve the quality of the quasi-ideal dataset.

\begin{figure}
\begin{center}
\includegraphics[width=0.50\textwidth]{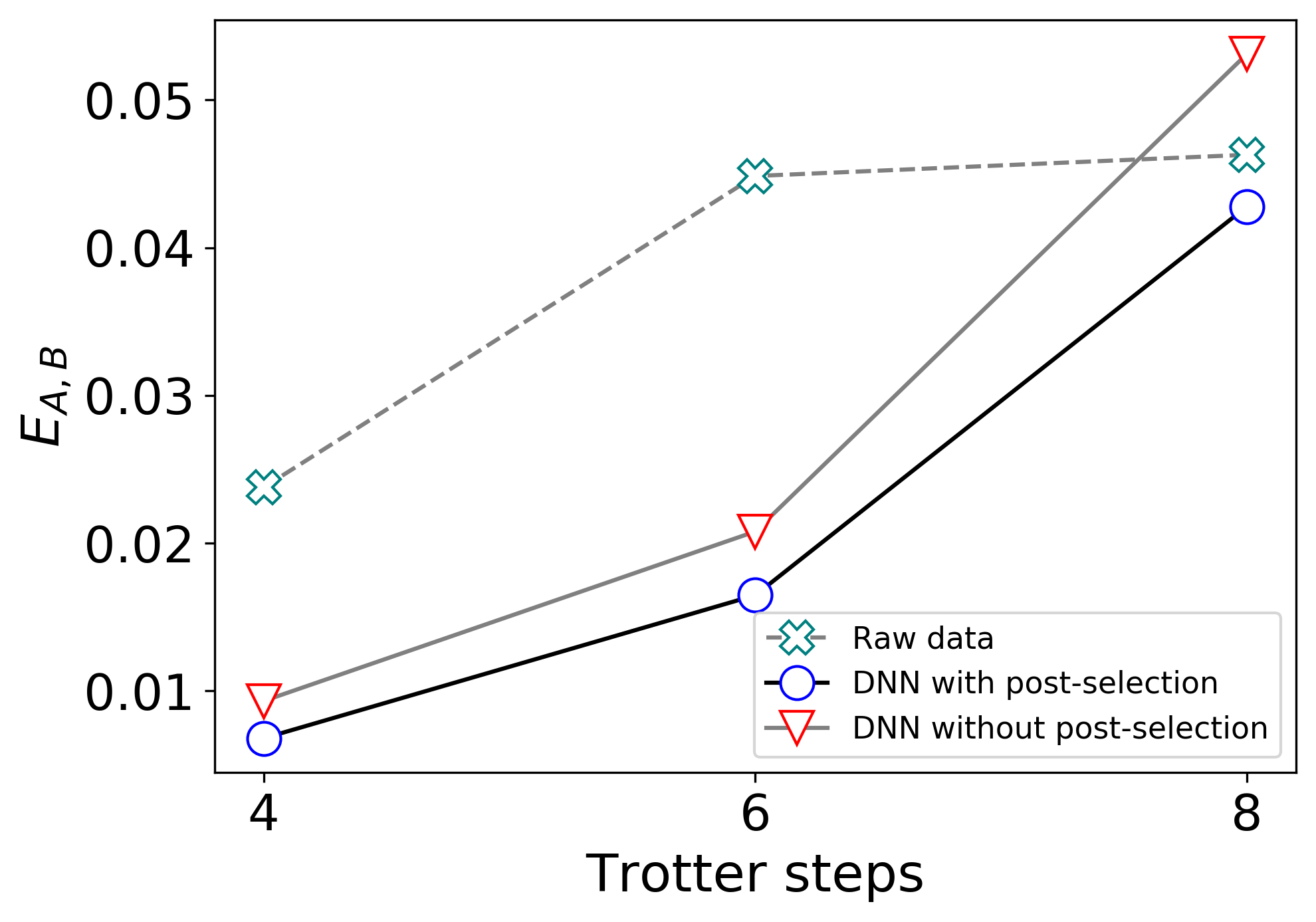} 
\caption{\label{fig:fig_mse_xy}MSE between ideal simulation data for a given Trotter step number and experimental data from an IBM Bogota 5-qubit quantum processor improved by the network with post-selection (o-shape symbols) and without post-selection (triangle-shape symbols) as well as raw data (x-shape symbols and dashed line). The errors were averaged over time and all initial states from the computational basis.}
\end{center}
\end{figure}

Figure \ref{fig:fig_mse_xy} shows MSE between ideal simulation data for a given Trotter step number at $k=z$ in Eq. (\ref{MSE}) and experimental data improved by the network with post-selection (o-shape symbols) and without post-selection (triangle-shape symbols) as well as raw data (x-shape symbols and dashed line). It is seen from this figure that neural network is able to significantly improve the quality of the data by decreasing MSE in several times for $N_2=4$ and 6. Post-selection leads to a further improvement.

\begin{figure}
\begin{center}
\includegraphics[width=0.99\textwidth]{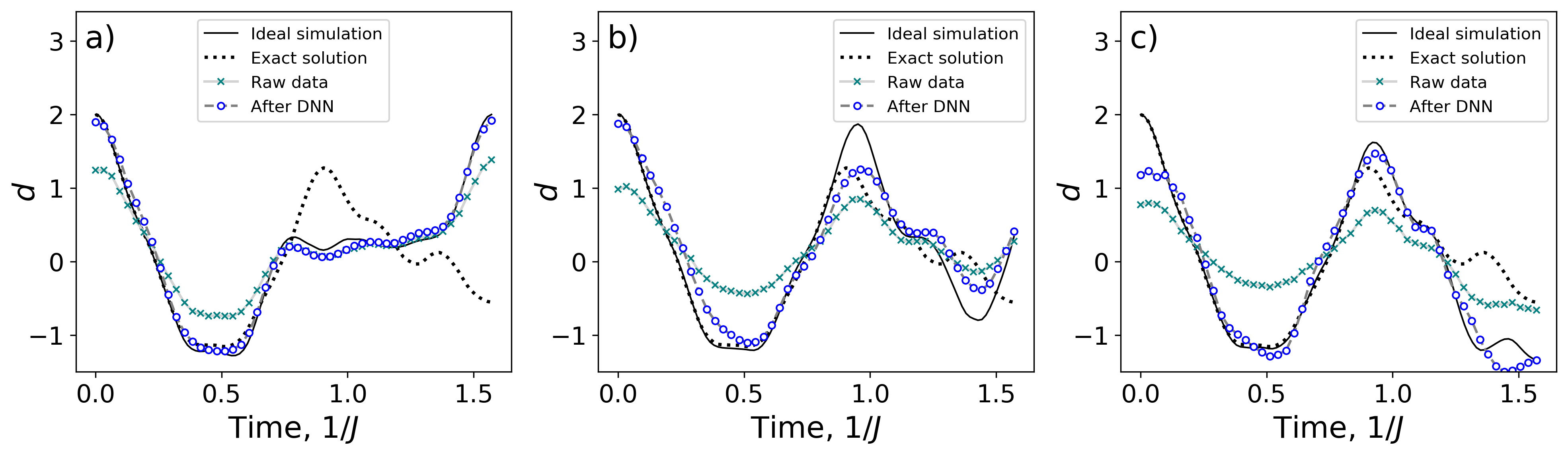} 
\caption{\label{fig:fig_xy}
The half-difference of spin magnetization along the $z$ axis as a function of time for the initial state $\ket{11100}$. Solid curves are the results of ideal simulations for given Trotter steps number $N_2$:  4 (a), 6 (b), 8 (c). Experimental data for half-difference of spin magnetization (x-shape symbols) were obtained on the real 5-qubit quantum processor IBM Bogota. The results of applying the neural network are shown with o-shape symbols. For each case, the neural network was trained using experimental data for $N_1=2$ and $c=2,3,4$, respectively (see the text). The dotted line shows the exact reference evolution of half-difference of spin magnetization free of Trotterization errors. In all cases $J=2$, $h=1$.}
\end{center}
\end{figure}

In order to illustrate the dynamics, we focus on the initial conditions, which are characterized by the domain wall between polarized and unpolarized spins, such as $\ket{11100}$, and consider a quantity known as half-difference of spin magnetization defined as
\begin{equation}
    d = \sum_{j \in S^{init}_{up}}\frac{1}{N^{init}_{up}}m_j - \sum_{j \in S^{init}_{down}}\frac{1}{N^{init}_{down}}m_j,
\end{equation}
where $m_j=2n_{j}-1$ is $j$-th spin magnetization, $S^{init}_{up}$ and $S^{init}_{down}$ are sets of spins with initial state $\ket{1}$ and $\ket{0}$, while $N^{init}_{down}$ and $N^{init}_{up}$ are their numbers, correspondingly.
The half-difference of spin magnetization
 characterises the robustness of theinitial spin pattern (the domain wall in our case) to the interaction between spins during the free evolution. The dynamics of this quantity is shown in Fig.~\ref{fig:fig_xy} for the initial state $\ket{11100}$ and different $N_2$.  In this particular case, e.g., $S^{init}_{up}=\{1,2,3\}$, $S^{init}_{down}=\{4,5\}$, and $N^{init}_{up}=3$, $N^{init}_{down}=2$. The post-selection technique was used at the training stage. From this figure, we see an improvement of the data quality due to the neural network.

\begin{figure}
\begin{center}
\includegraphics[width=0.99\textwidth]{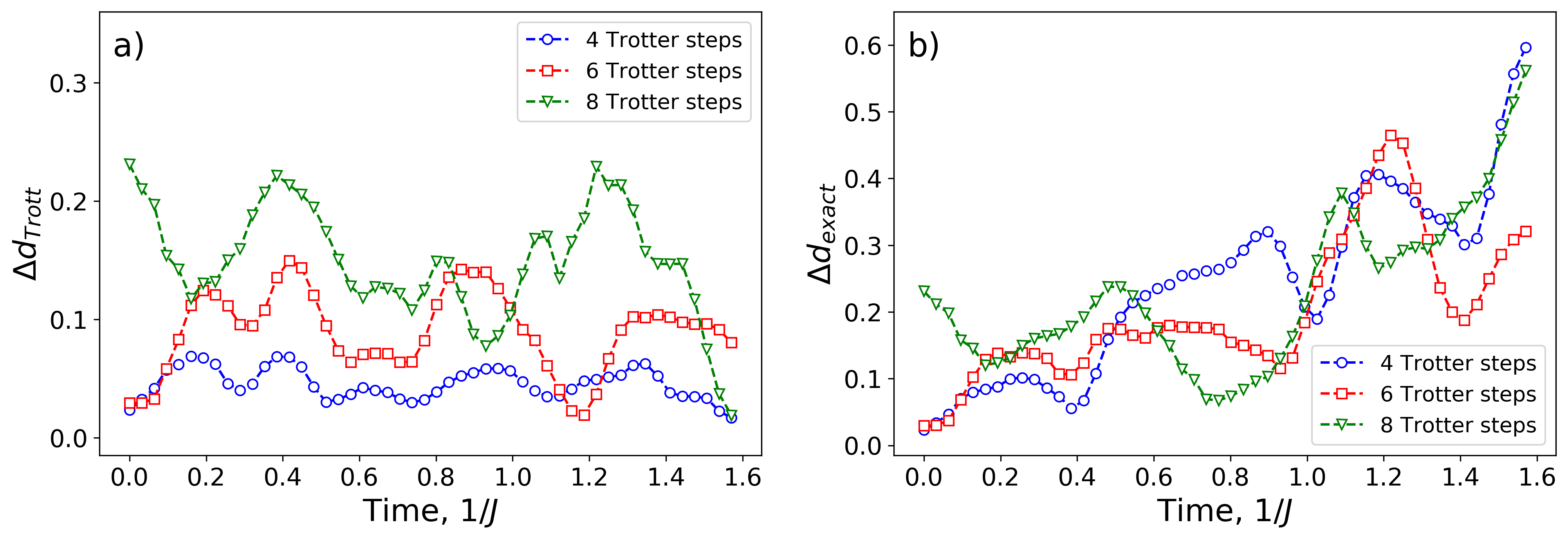} 
\caption{\label{fig:fig_errors_xy}
Errors after DNN application as a function of time for the initial state $\ket{11100}$ -  (a) difference between improved results and ideal simulation for the same Trotter step number;  (b) difference between improved results and exact results free of Trotterization error.}
\end{center}
\end{figure}

Figure~\ref{fig:fig_errors_xy} shows deviations of the improved results $\langle d \rangle_{DNN}$ from the ideal results with the same Trotter step number $\langle d \rangle_{Trott}$, $\Delta d_{Trott} = \langle d \rangle_{DNN}-\langle d \rangle_{Trott}$ (a) as well as of the results of the full solution of Schrodinger equation $\langle d \rangle_{exact}$ $\Delta d_{exact}=\langle d \rangle_{DNN}-\langle d \rangle_{exact}$ (b) as a function of time for the initial state $\ket{11100}$. There is an improvement of the dynamics for all values of $N_2$ considered. The comparison with the exact result free of trotterization errors presented in Fig. \ref{fig:fig_errors_xy}(b) shows that the deviation from this result generally grows with time for all values of $N_2$ considered that is qualitatively similar to the behavior revealed for the transverse-field Ising chain depicted in Fig. \ref{fig:fig_errors}(b).

\section{Conclusions}

In this article, we have proposed a method for the application of classical neural networks for the improvement of the outcomes of noisy quantum computers at the post-processing stage. In contrast to other suggestions, using our approach, it is possible to get data for training a neural network without relying on a classical simulator or any other source of ideal data. The approach is most suitable to the very important problem of digital simulation of quantum dynamics, for which a limitation in accessible Trotter steps numbers is currently crucial. The limitation is associated with the error accumulation due to the hardware imperfections.

Our method is based on artificial noise enhancement on the training stage that can be done by incorporation of fictitious Trotter blocks formally equivalent to identity gates into the circuit. Their role is to increase noise level due to the hardware imperfections while preserving the circuit's general structure and its relevant features. The network is trained to transform data obtained with such fictitious steps towards data obtained without them, that is, for rather shallow circuits, for which hardware errors are not critical. This idea seems to be more prospective for near-term generations of quantum computers with reduced gate errors, for which circuits at the training stage can already support large entanglement.

After being trained, the network can be applied to new data with the same Trotter step number, i.e., increased in the same way as at the training stage, but without fictitious Trotter steps. The amount of noise in this case is similar to that at the training stage. This trick allows for the effective increase of the Trotter number due to the post-processing, in the sense that errors become suppressed and results of simulations, which must have error rates below a given level, start to include data with larger Trotter step number.

Our method does not require a complete tomography of quantum states, which allows it to be scaled. The reason is that the network is trained to improve the data for a restricted number of quantum mean values such as spins magnetizations along different axes, order parameters, or characteristic correlators.

We have demonstrated the basic ingredients of our approach using two examples:  digital quantum simulations of the dynamics of the transverse-field Ising chain and XY chain. Deep neural network with simple architectures were used at the post-processing stage. For XY chain, an additional post-selection of the results at the training stage was applied by discarding a part of the data, which does not conserve the excitation number, as required by the Hamiltonian. The proof-of-principle results obtained on a real 5-qubit IBM Athens and Bogota quantum processors show that our method allows us to increase the number of Trotter steps while maintaining the same level of errors. The significant error reduction is the main result of our demonstration. A single neural network is able to improve the data for different initial conditions.

We believe that the proposed approach can be useful in the context of error mitigation in noisy quantum devices (especially of next generations with hardware errors decreased and qubit number increased). Particularly, it can be used in the case of periodic quantum circuits and in combination with other error reduction tools, such as post-selection or partial error correction.

\section*{Acknowledgements}
We acknowledge use of the IBM Quantum Experience for this work. The viewpoints expressed are those of the authors and do not reflect the official policy or position of IBM or the IBM Quantum Experience team.

A. A. Zh. acknowledges a support from RFBR (project no. 20-37-70028). W. V. P. acknowledges a support from RFBR (project no. 19-02-00421).

\backmatter

\bibliography{neuro}

\end{document}